\documentclass[pra,aps,twocolumn,nopacs,superscriptaddress,nofootinbib]{revtex4-1}

\usepackage{amsmath}  \usepackage{amssymb}  \usepackage{amsfonts}  \usepackage{bm}  \usepackage{bbm}  \usepackage{bbold}   \usepackage{braket}  \usepackage{color}  \usepackage{comment}  \usepackage{dcolumn}  \usepackage{enumerate}  \usepackage{epsfig}  \usepackage{gensymb}  \usepackage{graphicx}  \usepackage{indentfirst}  \usepackage{lmodern}  \usepackage{mathrsfs}  \usepackage{mathtools}  \usepackage{psfrag}  \usepackage{pst-all}  \usepackage{xcolor}
\usepackage{float} 
\usepackage[colorlinks,linkcolor=blue,citecolor=blue,urlcolor=blue,hyperindex,driverfallback=dvipdfm]{hyperref}  \usepackage[T1]{fontenc} 

\def\ii{{\rm i}}  \def\ee{{\rm e}}
\def\me{m_{\rm e}}  
  
        \def\Eb{{\bf E}}      \def\fb{{\bf f}}    \def\gb{{\bf g}}            \def\kb{{\bf k}}          \def\Qb{{\bf Q}}    \def\Rb{{\bf R}}  \def\rb{{\bf r}}      \def\vb{{\bf v}} 
\def\xx{\hat{\bf x}}    \def\zz{\hat{\bf z}}            
        
\def\Rmax{{R_{\rm max}}}  \def\kk{\hat{\bf k}}

\begin{document} 
\title{Spectrometer-Free Electron Spectromicroscopy}

\author{F.~Javier~Garc\'{\i}a~de~Abajo}
\email{javier.garciadeabajo@nanophotonics.es} 
\affiliation{ICFO-Institut de Ciencies Fotoniques, The Barcelona Institute of Science and Technology, 08860 Castelldefels (Barcelona), Spain} 
\affiliation{ICREA-Instituci\'o Catalana de Recerca i Estudis Avan\c{c}ats, Passeig Llu\'{\i}s Companys 23, 08010 Barcelona, Spain}
\author{Cruz~I.~Velasco}
\affiliation{ICFO-Institut de Ciencies Fotoniques, The Barcelona Institute of Science and Technology, 08860 Castelldefels (Barcelona), Spain} 

\begin{abstract}
We introduce an approach for performing spectrally resolved electron microscopy without the need for an electron spectrometer. The method involves an electron beam prepared as a coherent superposition of multiple paths, one of which passes near a laser-irradiated specimen. These paths are subsequently recombined, and their interference is measured as a function of laser frequency and beam position. Electron--light scattering introduces inelastic components into the interacting path, thereby disturbing the interference pattern. We implement this concept using two masks placed at conjugate image planes. The masks are complementary and act in tandem to fully suppress electron transmission in the absence of a specimen. However, electron interaction with an illuminated specimen perturbs the imaging condition, enabling electron transmission through the system. For a fixed external light intensity, the transmitted electron current is proportional to the strength of the local optical response in the material. The proposed technique does not require monochromatic electron beams, dramatically simplifying the design of spectrally resolved electron microscopes.
\end{abstract}

\maketitle 
\date{\today} 


\section{Introduction}

Since the invention of the electron microscope nearly a century ago \cite{R1987_2}, a sustained series of technical advances have permitted the generation of electron beams (e-beams) with a high degree of lateral coherence \cite{BDK02,NCD04,MKM08} and small energy broadening \cite{KUB09,KLD14}. State-of-the-art instruments allow electron energy-loss spectroscopy (EELS) to be performed with a combined spatiotemporal resolution in the sub-{\aa}ngstr\"om and few-meV regime \cite{KS14,HNY18}. Examples of application are the study of vibrational features with atomic detail \cite{HRK20}, the characterization of narrow excitons in transition-metal dichalcogenides \cite{SWW22}, and the identification of photonic modes in dielectric cavities \cite{paper383}. However, EELS is still unable to resolve narrow spectral features associated with high-quality-factor optical excitations of interest for nanophotonics, such as long-lived excitons and narrow resonances in dielectric cavities, which are highly dependent on the nanostructured material environment \cite{WT24,paper441}.

Electron energy-gain spectroscopy (EEGS) has emerged as a unique technique capable of reaching a spectral resolution in the $\mu$eV range by recording the fraction of electrons whose energy is increased when traversing a laser-irradiated specimen \cite{paper114,paper221,LWH19,HRF21,paper419}. In EEGS, the electron interacts with the optical field scattered by the specimen, which can contain evanescent components with sufficient momentum to break their kinematic mismatch with photons in free space, thus producing inelastic electron--light scattering (IELS). The probability that electrons absorb photons is proportional to the optical near-field intensity, which is boosted when the laser frequency matches optical resonances in the specimen. Consequently, even with a modest electron energy resolution in the sub-eV range, a high spectral resolution can be achieved by fine-tuning the photon energy \cite{paper114,HRF21,paper419}. In a related context, although cathodoluminescence (CL) spectroscopy can also yield a high spectral resolution, it is limited by relatively weak signals, especially when looking at narrow resonances \cite{paper419}.

Soon after EEGS was proposed \cite{paper114}, a temporal resolution in the femtosecond domain was demonstrated in electron microscopy through the synthesis of synchronized femtosecond electron and laser pulses in the so-called photon-induced near-field electron microscopy (PINEM) technique \cite{BFZ09,paper151,FES15,PLQ15,RB16,PRY17,paper332,KLS20,DNS20,HRF21,NKS23,BNS24,paper431}. Early attempts to perform EEGS based on PINEM \cite{WDS20,KDS21} produced an energy resolution similar to EELS and limited by the short duration of the electron pulses due to the uncertainty principle. To push energy resolution to the sub- or few-$\mu$eV range, continuous-wave \cite{HRF21} or nanosecond \cite{paper419} lasers were necessary.

The techniques mentioned above require electron spectrometers and monochromators. For example, in EEGS and PINEM, electron spectra need to be collected with better resolution than the employed photon energy. This condition was relaxed in the method introduced in Ref.~\cite{paper306}, where a broad energy filtering was sufficient to resolve photon-induced changes in the e-beam. The ability to collect spatially resolved spectral information on a specimen without using an electron spectrometer could simultaneously eliminate the need for electron monochromators and make the acquisition of spectral information more widely available in electron microscopes.

\begin{figure}
\begin{centering} \includegraphics[width=0.43\textwidth]{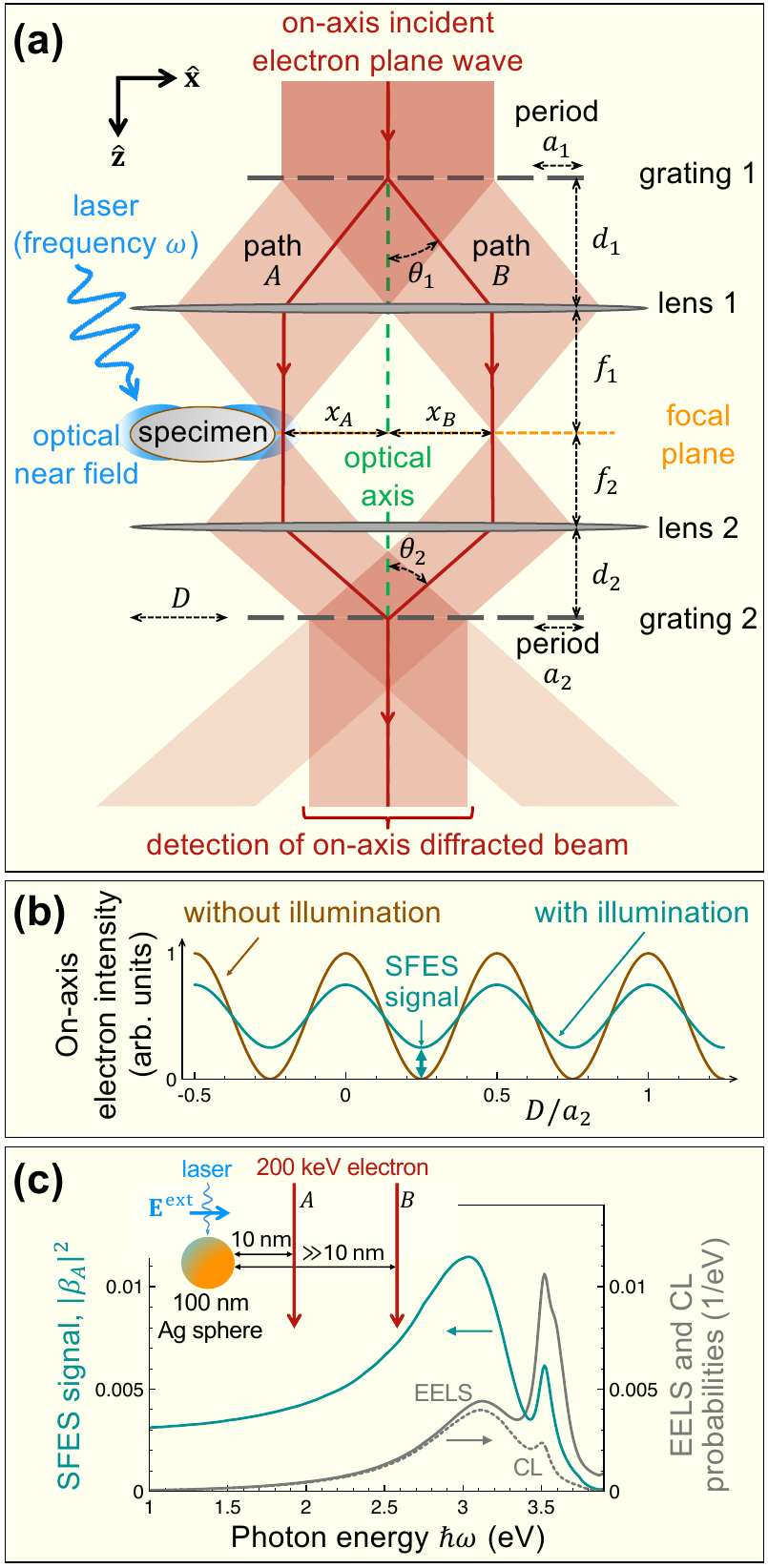} \par\end{centering}
\caption{Principle of spectrometer-free electron spectromicroscopy (SFES). (a)~An incident e-beam is prepared as an on-axis electron plane wave, which is diffracted by a periodic grating 1 (period $a_1$ along $\xx$). We consider two dominant diffracted beams (paths $n=A,B$) forming an angle $\theta_{1}$ relative to the optical axis $\zz$.
A lens 1 (focal length $f_1$) focuses these beams on the sample plane at lateral positions $\Rb_n=x_n\xx$. A second lens 2 (focal length $f_2$) transforms each beam $n$ into a plane wave that is in turn diffracted by a second grating 2 (period $a_2$, lateral displacement $D$ along $\xx$). Path $A$ is closest to the specimen and interacts with the scattered optical field upon illumination by a laser of frequency $\omega$. (b)~The transmitted electron intensity in the on-axis diffracted beam displays an interference pattern as a function of $D$, whose visibility is modulated by IELS via the depletion of the elastic channel depending on the optical near-field strength at the specimen. We adjust $D$ to suppress on-axis transmission without illumination. (c)~An SFES spectrum is constructed as a function of illumination frequency $\omega$ (left scale, with an external field $E^{\rm ext}=2$~MV/m) similar to those of EELS and CL (right scale, for a single beam along path $A$), here presented for the geometry and specimen sketched in the inset.}
\label{Fig1}
\end{figure}

In this article, we introduce a method to perform spectrometer-free electron spectromicroscopy (SFES) with spectral and spatial resolution comparable to EEGS. The principle of SFES is depicted in Fig.~\ref{Fig1}(a): an electron plane wave is split into two beams ($A$ and $B$) that are focused in the plane of the specimen and later recombined into a single e-beam, where they interfere; the visibility of the interference pattern constitutes the measured signal; the specimen is illuminated by a laser that produces IELS analogous to PINEM on the electron path $A$ passing closest to it; the interference is thus depleted (in proportion to the optical near-field intensity) by subtracting electrons from the elastic channel in path $A$, while path $B$ remains nearly elastic because it passes away from the specimen. We present a detailed analysis of the SFES electron signal expected under attainable conditions, along with an extension to multiple-path splitting that should facilitate the implementation of SFES by eliminating the need for angular resolution (i.e., all transmitted electrons are detected without resolving them in angle).

\section{Tutorial implementation of SFES}

The incident two-path electron state considered in Fig.~\ref{Fig1}(a) can be written as $\ket{\Psi^{\rm inc}}=\big(\ket{A}+\ket{B}\big)\otimes\ket{0}$ before interaction with the specimen, where $\ket{A}$ and $\ket{B}$ denote the mutually orthogonal transverse part [along $\Rb=(x,y)$ directions] associated with the respective paths while $\ket{0}$ refers to the initial longitudinal component (along the e-beam direction, $z$). Following a well-established theoretical description of PINEM \cite{paper151,PLZ10,paper371}, the electron state after interaction with the illuminated specimen can be written as
\begin{subequations}
\begin{align} \label{Psi} 
\ket{\Psi}=\sum_\ell\big(\alpha_{A,\ell}\ket{A}+\alpha_{B,\ell}\ket{B}\big)\otimes\ket{\ell},
\end{align}
where $\ket{\ell}$ is the longitudinal electron component after exchanging $\ell$ photons with the optical field, while the corresponding IELS amplitudes
\begin{align} \label{alphan} 
\alpha_{n,\ell}=J_\ell(2|\beta_n|)\ee^{\ii\ell{\rm arg}\{-\beta_n\}}
\end{align}
depend on the coupling coefficients
\begin{align} \label{betan} 
\beta_n=(e/\hbar\omega)\int dz E_z(\Rb_n,z)\ee^{-\ii\omega z/v}
\end{align}
\end{subequations}
determined by the spatial Fourier transform of the $z$ component of the optical electric field $\Eb(\rb)$ evaluated along the paths $n=A$ and $B$ crossing the focal points $\Rb_n$ in the sample plane. Here, we consider a monochromatic, coherent optical field $\Eb(\rb,t)=\Eb(\rb)\ee^{-\ii\omega t}+{\rm c.c.}$ of frequency $\omega$, which is assumed to vary negligibly across the width of each focused beam. In addition, the transverse electron state is assumed to be unaffected by the interaction under the nonrecoil approximation \cite{paper149}.

For tutorial purposes, Fig.~\ref{Fig1} focuses on symmetric gratings transmitting only the first two diffraction orders, which correspond to in-plane wave vector transfers $\pm2\pi/a_i$, where $+$ and $-$ refer to paths $n=A$ and $B$, while $i=1,2$ denotes the grating index (periods $a_i$); also, we draw $f_i=d_i$ for clarity, but this condition is neither required nor assumed in what follows. The paraxial angles $\theta_i$ after and before diffraction by gratings 1 and 2, respectively, must be coordinated so that an on-axis beam component is obtained after the entire process. This condition is fulfilled by imposing the relation
\begin{align} \label{gratingconjugation} 
a_1/a_2=f_1/f_2
\end{align}
on the focal distances and periods in Fig.~\ref{Fig1}(a) (see Appendix\ \ref{conjugation}). We further assume path-independent transmission coefficients $t_i$ and note that a lateral displacement $D$ of grating 2 introduces a phase in these coefficients, which take the path-dependent values $\ee^{\mp2\pi\ii D/a_2}\,t_2$  (see Appendix\ \ref{displacement}). In the configuration of Fig.~\ref{Fig1}(a) without external illumination, the fraction of electrons transmitted along the on-axis beam becomes
\begin{align} \label{P0} 
P_0(D)=4|t_1t_2|^2\cos^2\big(2\pi D/a_2\big),
\end{align}
which displays an interference pattern as a function of $D$ [Fig.~\ref{Fig1}(b), without illumination]. The grating displacement $D$ can be adjusted to compensate for any uncontrolled phase difference between the two paths. In the presence of illumination, the superposition of different inelastic channels in the post-interaction wave function [Eq.~(\ref{Psi})] leads to a probability of on-axis transmission modified from Eq.~(\ref{P0}) to (see Appendix\ \ref{Gaussianspecimen})
\begin{align} 
P(D)=&P_0(D) \label{PP}\\
&+2|t_1t_2|^2{\rm Re}\Big\{\ee^{4\pi\ii D/a_2}\Big(\sum_\ell\alpha_{A,\ell}\alpha_{B,\ell}^*-1\Big)\Big\}, \nonumber
\end{align}
where the coefficients $t_i$ are assumed to vary negligibly within the range of inelastic energy transfers under consideration, and we have used the property $\sum_\ell|\alpha_{n,\ell}|^2=1$ (i.e., the interaction with light does not change the total electron probability).

Taking path $B$ sufficiently far away from the specimen to neglect its interaction with light, we have $\alpha_{B,\ell}=\delta_{\ell0}$, and consequently, Eq.~(\ref{PP}) becomes $P(D)=P_0(D)+2|t_1t_2|^2\cos\big(4\pi D/a_2\big)\big[J_0(2|\beta_A|)-1\big]$. We are interested in the difference produced in the electron current with and without illumination [SFES signal, see Fig.~\ref{Fig1}(b)]. One possibility would be to record this current as a function of the displacement $D$. A more convenient approach is adjusting $D$ by measuring the electron current without a specimen to suppress the on-axis transmission when $D=D_0=a_2/4+ma_2/2$ for any integer $m$. Then, the SFES probability of electron transmitting along the on-axis beam reduces to
\begin{align} \label{PSFES} 
P(D_0)=2|t_1t_2|^2\big[1-J_0(2|\beta_A|)\big]\approx2|t_1t_2|^2|\beta_A|^2,
\end{align}
where the rightmost approximation applies to small IELS coefficients $|\beta_A|\ll1$.

We plot $|\beta_A|^2$ in Fig.~\ref{Fig1}(c) for a configuration in which path $A$ passes near a silver sphere featuring a dipolar mode (prominent broad peak) as well as higher-order resonances. The result is compared to EELS and (angle-integrated) CL spectra for an electron following a single path $A$ (see Appendix\ \ref{simulations} for details of the calculation). The SFES spectrum is similar to that obtained with CL, as expected from the fact that the electron--light coupling coefficient $\beta$ and the CL emission amplitude $\fb$ are related through the rigorous relation \cite{paper371} $\beta=\ii c^2 \fb\cdot\Eb^{\rm ext}/(\hbar\omega^2)$, provided $\beta$ refers to an electron moving along a trajectory $\rb_0+\vb t$ in combination with external plane-wave illumination of field amplitude $\Eb^{\rm ext}$ and wave vector $\kb=(\omega/c)\,\kk$, while $\fb$ corresponds to emission with wave vector $-\kb$ produced by an electron moving along the trajectory $\rb_0-\vb t$. In contrast, EELS does not require far-field coupling, thus placing a comparatively higher weight on nondipolar modes.

\begin{figure}
\begin{centering} \includegraphics[width=0.43\textwidth]{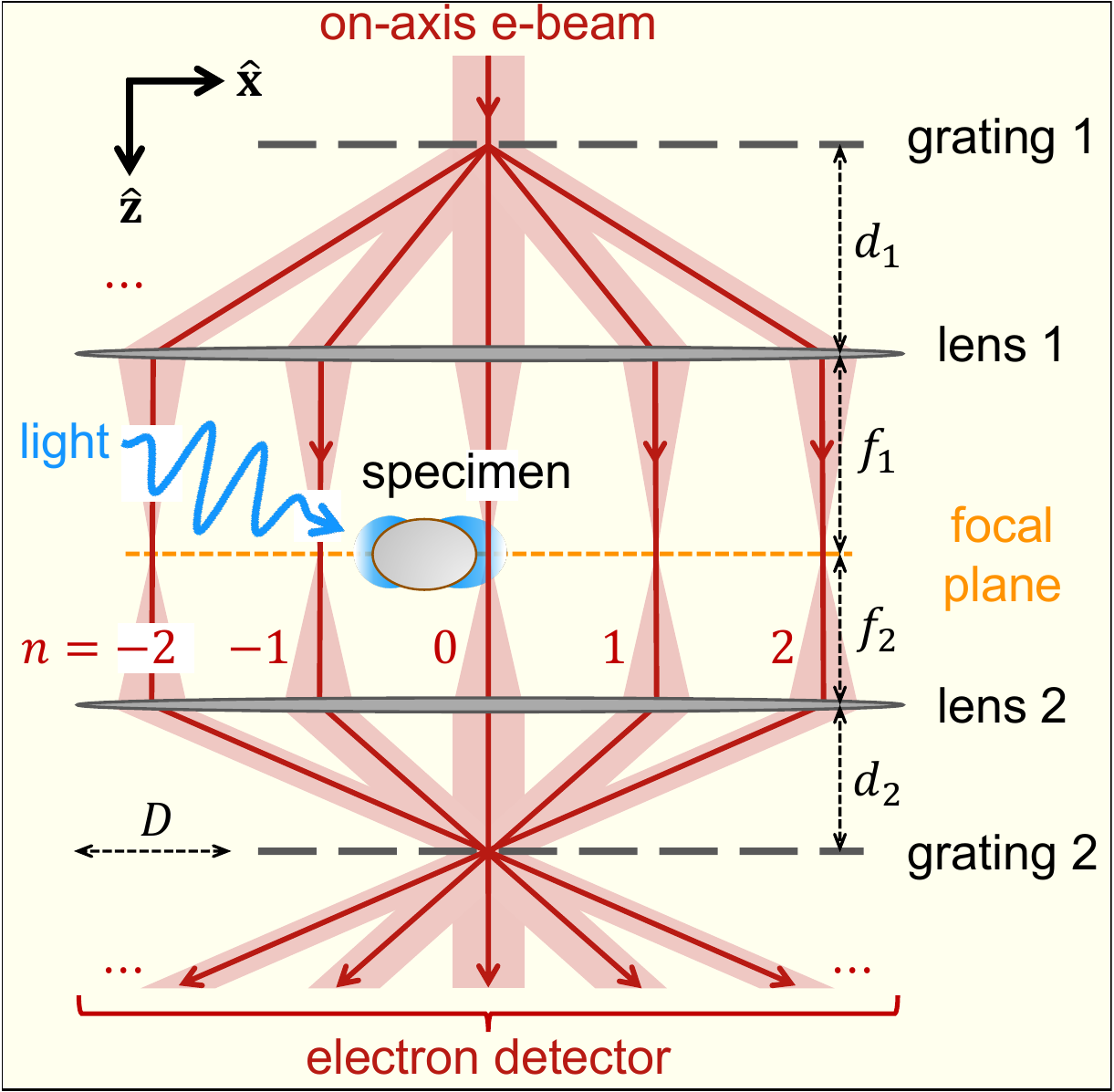} \par\end{centering}
\caption{SFES based on the detection of all diffracted beams. We consider an alternative scheme to that shown in Fig.~\ref{Fig1}(a) to eliminate the need for angular resolution by collecting all electron diffraction orders.}
\label{Fig2}
\end{figure}

\section{SFES based on complementary conjugated masks}

While the scheme in Fig.~\ref{Fig1}(a) is convenient for explaining the principle of SFES, it requires angle-resolved electron detection to filter the on-axis transmitted e-beam, which can be challenging for the small diffraction angles typically produced by transmission electron gratings \cite{paper388} (e.g., first-order beams are transmitted with an off-axis angle $\theta_2=\sin^{-1}(2\pi\hbar/\me a_2v)\approx17\,\mu$rad for $a_2=200$~nm and 200~keV electrons). A more convenient configuration that eliminates the need for angular resolution is presented in Fig.~\ref{Fig2}, where all diffracted beams are collected by the electron detector and grating coordination [Eq.~(\ref{gratingconjugation})] is still assumed. We further impose the condition $f_1^2/f_2^2=-(d_1-f_1)/(d_2-f_2)$, which guarantees that each grating lies in the image plane of the other one through the two-lens system. Because the electron wavelength is small compared to the periods, we can describe the gratings through local periodic transmission functions $t_j(x)=t_j(x+a_j)$, whose Fourier transforms $t_{j,n}=\int_{-a_j/2}^{a_j/2} dx\,t_j(x)\,\ee^{-2\pi\ii nx/a_2}$ give the amplitudes of the diffracted beams. In the absence of a specimen, the probability of electron transmission through the setup is given by (see Appendix\ \ref{Gaussianspecimen} for a detailed derivation)
\begin{align}
P_0(D)=\int_{-1/2}^{1/2}d\theta\;\big|t_1(\theta a_1)t_2(-\theta a_2-D)\big|^2, \label{SFES4}
\end{align}
which, as expected, corresponds to the superposition of the two gratings.

We are interested in suppressing electron transmission in the absence of a specimen, such that the SFES signal associated with the interaction between the electrons and an illuminated specimen is not contaminated by a dark background. With black-and-white gratings [i.e., $t_j(x)\in\{0,1\}$] such as slit arrays, suppression is possible if the openings cover (at most) half of the total surface area. For example, when the slit widths are half of the periods [e.g., $t_j(x)=\Theta(a_j/4-x)$ for $|x|<a_j/2$], the probability, which is periodic with period $a_2$ [i.e., $P_0(D)=P_0(D+a_2)$], becomes $P_0(D)=1/2-|D|/a_2$ in the first period (i.e., for $|D|<a_2/2$) and, therefore, features a complete depletion of electron transmission for $D=a_2/2$. By making the slit openings narrower, Eq.~(\ref{SFES4}) predicts that complete suppression of transmission occurs over a finite range of $D$ values, which should render SFES more tolerant to an imperfect lateral coherence of the e-beam.

\begin{figure*}
\begin{centering} \includegraphics[width=0.95\textwidth]{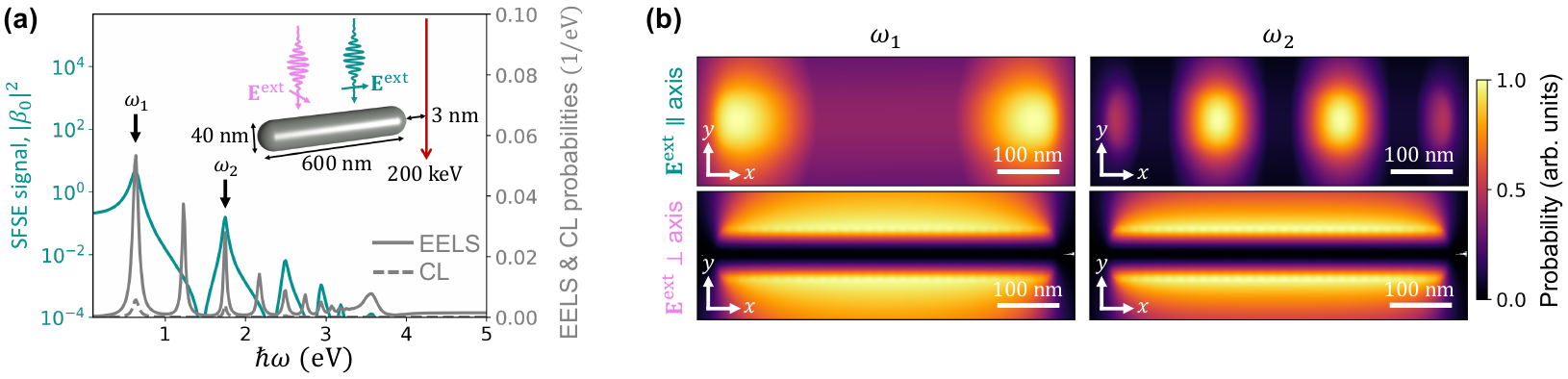} \par\end{centering}
\caption{SFES mapping of localized optical excitations. (a)~Comparison of EELS and CL spectra with the electron--light coupling strength $|\beta_0|^2$ in SFES for a silver nanowire under the geometry depicted in the inset. In SFES, we consider illumination with an incident electric field amplitude of 1~MV/m ($\sim5$~GW/m$^2$ intensity) and polarization parallel to the nanowire axis. (b)~SFES maps as a function of e-beam position for light polarization either parallel (top) or perpendicular (bottom) to the nanowire axis and frequencies $\omega_1$ and $\omega_2$ tuned to the first dipole-active modes of the particle.}
\label{Fig3}
\end{figure*}

When placing an illuminated specimen at the focal plane in the scheme of Fig.~\ref{Fig2}, the electron transmission probability becomes (see Appendix\ \ref{Gaussian})
\begin{align}
P(D)=\sum_{n,\ell}\bigg|\sum_{n'} \alpha_{n',\ell} \;t_{1,n'} \;t_{2,n'+n} \;\ee^{-\ii n'g_2D}\bigg|^2, \label{SFES5}
\end{align}
where we have introduced the IELS amplitudes $\alpha_{n',\ell}$ defined in Eq.~(\ref{alphan}). Different $n'$ beams diffracted by grating 1 undergo interference within each final electron energy subspace $\ell$, while different $n$ beams (after diffraction by grating 2) are summed incoherently. For simplicity, we consider that only the $n'=0$ focused beam interacts with the optical field so that the rest of the beams pass away from the specimen and have amplitudes $\alpha_{n'\neq0,\ell}=\delta_{\ell,0}$ (i.e., no inelastic components). Further setting the grating configuration (including $D=D_0$) such that $P_0=0$, Eq.~(\ref{SFES5}) reduces to (see Appendix\ \ref{Gaussian})
\begin{align}
P(D_0)=2\,|t_{10}|^2 \;T_2 \;\big[1-J_0(2|\beta_0|)\big], \label{SFES6}
\end{align}
where $T_2=\sum_n|t_{2,n}|^2$. For example, for slit-based gratings with slit widths $a_j/2$, we have $|t_{10}|^2=1/4$ and $T_2=1/2$, and considering a coupling coefficient $|\beta_0|\ll1$, the probability becomes $P\approx|\beta_0|^2/4$.

In a feasible implementation, grating periods $a_j$ of 100's~nm \cite{paper388} could be used in combination with $f_j$ and $d_j$ distances in the millimeter range, thus rending a linear periodic array of foci with a period $g_1f_1/q_0$ of 10's~nm. This should be enough to place the central $n'=0$ focused beam near the specimen while the rest of the beams are passing far from it.

For illustration, Fig.~\ref{Fig3} shows the mapping of plasmons in a long silver nanowire via SFES. Like in the sphere of Fig.~\ref{Fig1}, the SFES spectrum (i.e., $|\beta_0|^2$ as a function of light frequency) yields similar information to EELS [Fig.~\ref{Fig3}(a)]. However, the near-field enhancement is larger in the nanowire, thus producing stronger IELS coupling for a given incident light intensity. By raster scanning the central beam and tuning the light frequencies to the first dipole-active modes, we observe the characteristic maps shown in the top panels of Fig.~\ref{Fig3}(b). Like EEGS, SFES allows us to select the external light polarization to show that these modes are polarized along the wire (i.e., featureless maps are obtained under transverse polarization, as shown in the bottom panels).

\section{Concluding remarks}

In summary, our proposed scheme can yield spectrally and spatially resolved information by measuring the transmitted free-electron current without needing either electron spectrometers or angle-resolved detection. A spectral resolution in the (sub-)$\mu$eV range should be attainable with nanosecond laser pulses illuminating the sample in combination with electron pulses of similar duration using already demonstrated EEGS setups \cite{paper325,paper419}. This level of temporal precision can be achieved through synchronized electron blanking like in EEGS \cite{paper419}, either using electrostatic e-beam deflection of the recently developed microwave-cavity electron pulsing \cite{DFL23,BDV24}. This method is compatible with the use of spectrally broad e-beams, whose temporal coherence becomes irrelevant. Additionally, only a moderate level of e-beam collimation is required, so that the lateral coherence length of the incident electrons lies in the micron range, covering a sufficient number of grating periods to produce well-defined diffracted (and focused) beams. By imaging two overlapping gratings on each other, the condition of total e-beam depletion in the absence of an illuminated specimen is robust (e.g., tolerant to small grating misalignments and e-beam quality), and therefore, permitting the SFES signal to be unaffected by a dark background current. The SFES signal relies on stimulated IELS, and thus, it can be modulated by varying the light intensity, in contrast to the spontaneous processes probed by EELS and CL. In addition, optical phase information could be obtained through homodyne detection by exposure of the e-beam to optical near fields before interaction with the illuminated specimen \cite{paper431}. Both elastic scattering by the specimen and spontaneous inelastic interactions can impact the electron depletion condition in thick samples, which should require further analysis for the extraction of near-field information from SFES. As an extension of the proposed scheme, conjugated masks with more general patterns could be used instead of gratings to prepare, for example, vortex beams \cite{VTS10,BSV12} and still impose the condition of e-beam depletion in the absence of a specimen. The interaction of multiple beams with illuminated regions of the specimen could also be conceived, which would require further data analysis to resolve the contribution of the near field at each focal spot.

\section*{Acknowledgments}

This work was supported by the European Research Council (Grant No. 101141220-QUEFES).

\appendix

\begin{widetext}

\section*{Appendix}

We consider an e-beam propagating along the positive $z$ direction under paraxial conditions. The incident electrons have a kinetic energy $\hbar\varepsilon_0$, velocity $v=c\,\sqrt{1-1/(\hbar\varepsilon_0/\me c^2+1)^2}$, and longitudinal wave vector $q_0=\me v\gamma/\hbar$, where $\gamma=1/\sqrt{1-v^2/c^2}$ \cite{J99}. We further assume that any changes produced in the e-beam by gratings, lenses, and inelastic electron--light scattering (IELS) result in negligible fractional variations of these parameters (nonrecoil approximation). Without loss of generality, it is convenient to factorize the electron wave function as $\psi(\rb,t)=\ee^{\ii(q_0z-\varepsilon_0t)}\phi(\rb,t)$, where $\phi(\rb,t)$ is a slowly varying envelope that we shall use instead of $\psi(\rb,t)$ in what follows. In Secs.~\ref{freelectron}--\ref{fullSFES}, we study plane waves, lenses, and gratings with infinite lateral extension for simplicity. In Secs.~\ref{Gaussian}--\ref{Gaussianspecimen}, realistic conditions are considered by combining the results of previous sections to describe Gaussian e-beams as a combination of plane waves, assuming a lateral beam width smaller than the lens and grating diameters.

\renewcommand{\theequation}{A\arabic{equation}} 
\renewcommand{\thesection}{A} 
\section{Free electron propagation}
\label{freelectron}

The free-space propagation of a free electron is described by the expression
\begin{align}
\phi(\Rb,z_1)&=\int \frac{d^2\Qb}{(2\pi)^2}\,\ee^{\ii\Qb\cdot\Rb+\ii(q_z-q_0)(z_1-z_0)}\int d^2\Rb' \;\ee^{-\ii\Qb\cdot\Rb'} \phi(\Rb',z_0) \nonumber\\
&\approx\frac{-\ii q_0}{2\pi\,(z_1-z_0)}\int d^2\Rb' \;\ee^{\ii q_0|\Rb-\Rb'|^2/2(z_1-z_0)} \phi(\Rb',z_0), \label{eq1}
\end{align}
which relates the wave function $\phi(\Rb,z_1)$ at a transverse plane $z=z_1$ to the wave function $\phi(\Rb',z_0)$ at $z=z_0$. The first line in Eq.~(\ref{eq1}) represents a concatenation of both direct and inverse Fourier transforms in the transverse real and wave-vector coordinates $\Rb=(x,y)$ and $\Qb=(q_x,q_y)$, respectively. We note that a propagation phase corresponding to the change in the longitudinal wave vector $q_z=\sqrt{q_0^2-Q^2}$ relative to $q_0$ is introduced in wave-vector space. The second line in Eq.~(\ref{eq1}) is derived by applying the paraxial approximation $q_z\approx q_0-Q^2/2q_0$ and integrating over $\Qb$.

For an incident plane wave with transverse momentum $\hbar\Qb_i$, we can substitute $\ee^{\ii\Qb_i\cdot\Rb'}$ for $\phi(\Rb',z_0)$ in Eq.~(\ref{eq1}) and perform the $\Rb'$ integral analytically to yield
\begin{align}
\phi(\Rb,z_1)\approx\ee^{\ii\Qb_i\cdot\Rb-\ii Q_i^2(z_1-z_0)/2q_0}, \label{eq2}
\end{align}
which trivially agrees with the paraxial approximation applied to the longitudinal dependence of the incident wave $\psi\propto\ee^{\ii q_{iz} z}\approx\ee^{\ii q_0z-\ii Q_i^2z/2q_0}$ with $q_{iz}=\sqrt{q_0^2-Q_i^2}$.

\renewcommand{\theequation}{B\arabic{equation}} 
\renewcommand{\thesection}{B} 
\section{Focusing by an electron lens}
\label{focusing}

The role of a lens is to imprint a position-dependent phase shift on the transmitted electron wave function. Neglecting aberrations and restricting our analysis to the paraxial regime, the phase introduced by an axially symmetric lens increases quadratically with the radial distance $R$ to the $z$ axis. More precisely, transmission through a lens with focal length $f$ transforms the electron wave function as follows:
\begin{align}
\phi(\rb) \rightarrow \phi(\rb)\,\ee^{-\ii q_0R^2/2f}. \label{lens}
\end{align}
Considering an incident plane wave $\ee^{\ii\Qb_i\cdot\Rb}$ with $Q_i\ll q_0$ and including the effect of propagation over a distance $z-z_0=f$ after passing through the lens, the wave function at a plane $z$ near the focal region is given by Eq.~(\ref{eq1}) with $\phi(\Rb',z_0)$ replaced by $\ee^{\ii\Qb_i\cdot\Rb'-\ii q_0R'^2/2f}$ inside the integrand:
\begin{align}
\phi(\rb)&\approx\frac{-\ii q_0}{2\pi f}\ee^{\ii q_0R^2/2f}\int d^2\Rb' \;\ee^{\ii(\Qb_i-q_0\Rb/f)\cdot\Rb'}. \nonumber
\end{align}
For a laterally extended plane wave passing through a finite-aperture lens of radius $\Rmax$, the $\Rb'$ integral can be performed analytically to yield $\phi(\Rb,z)=(-\ii q_0R^2_{\rm max}/f)\;\ee^{\ii q_0R^2/2f}\times{\rm PSF}(|\Rb-\Rb_i|q_0\Rmax/f)$, where $\Rb_i=f\Qb_i/q_0$ is the lateral position of the focused beam and ${\rm PSF}(\theta)=J_1(\theta)/\theta$ is the point spread function. The focused beam features a central peak with a FWHM given by $\approx0.51\,\lambda_e f/\Rmax$, where $\lambda_e=2\pi/q_0$ is the electron wavelength. However, electron lenses are commonly designed to have large diameters compared to the e-beam diameter, so all electrons are transmitted. Therefore, we can perform the $\Rb'$ integral without restriction, yielding 
\begin{align}
\phi(\rb)&\approx\frac{-2\pi\ii f}{q_0}\ee^{\ii q_0R_i^2/2f}\;\delta(\Rb-\Rb_i). \label{PSF}
\end{align}
It should be stressed that this expression is only meaningful after performing an average over $\Qb_i$ (or equivalently $\Rb_i$) to produce a finite transverse width of the incident e-beam (see Sec.~\ref{Gaussianspecimen} below).

\renewcommand{\theequation}{C\arabic{equation}} 
\renewcommand{\thesection}{C} 
\section{Transmission through two conjugated lenses}
\label{twolenses}

In SFES, we consider two lenses of focal lengths $f_1$ and $f_2$ separated by a total distance $f_1+f_2$, as shown in Fig.~\ref{Fig1}a of the main text. Digregarding the gratings and following an analysis analogous to that in Sec.~\ref{focusing}, an incident plane wave $\ee^{\ii\Qb_i\cdot\Rb'}$ must be first multiplied by a phase factor $\ee^{-\ii q_0R'^2/2f_1}$ to account for transmission through lens 1 [see Eq.~(\ref{lens})], then propagated along a distance $z_1-z_0=f_1+f_2$ using Eq.~(\ref{eq1}), and finally multiplied by $\ee^{-\ii q_0R^2/2f_2}$ to account for lens 2. From this procedure, the wave function right after transmission through lens 2 is found to be
\begin{align}
\phi(\rb)&\approx\frac{-\ii q_0}{2\pi\,(f_1+f_2)}\int d^2\Rb' \;\exp\bigg\{\frac{\ii q_0|\Rb-\Rb'|^2}{2(f_1+f_2)}+\ii\Qb_i\cdot\Rb'-\frac{\ii q_0R'^2}{2f_1}-\frac{\ii q_0R^2}{2f_2}\bigg\} \nonumber\\
&=-\frac{f_1}{f_2}\;\ee^{\ii\varphi}\;\ee^{-\ii(f_1/f_2)\,\Qb_i\cdot\Rb}, \label{twolenseseq}
\end{align}
where $\varphi=(Q_i^2/2q_0)(f_1+f_1^2/f_2)$ is a global phase. The incident plane wave is thus deflected in the opposite transverse direction with a wave vector scaled by a factor $f_1/f_2$ relative to $-\Qb_i$.

\renewcommand{\theequation}{D\arabic{equation}} 
\renewcommand{\thesection}{D} 
\section{Beam splitting and mixing by transmission gratings}
\label{grating}

A transmission grating periodic in $x$ with period $a$ can be described by a local transmission function $t(x)=t(x+a)$, so that an incident plane wave $\ee^{\ii\Qb_i\cdot\Rb}$ is transformed into $\ee^{\ii\Qb_i\cdot\Rb}\,t(x)$. The assumption of a local function $t(x)$ is justified under paraxial propagation conditions ($Q_i\ll q_0$), provided the period $a$ is large compared with the electron wavelength (e.g., $\lambda_e\approx2.5$~pm at 200~keV). In addition, for the large kinetic energies under consideration, electron reflection can be disregarded. We further ignore corrections introduced in $t(x)$ by inelastic and image interactions with the grating structure \cite{paper357} as well as radiative decoherence produced by material-assisted coupling to radiation \cite{paper425}.

It is convenient to Fourier-transform the transmission function as
\begin{subequations}
\label{gratingeq}
\begin{align}
&t(x)=\sum_n t_n\,\ee^{\ii ngx}, \label{gratingeq1}\\
&t_n=\frac{1}{a}\int_{-a/2}^{a/2} dx \;t(x)\,\ee^{-\ii ngx}, \label{gratingeq2}
\end{align}
\end{subequations}
where $g=2\pi/a$ is the unit vector of the reciprocal lattice and the integer index $n$ runs over diffraction orders with amplitudes $t_n$. Incidentally, the fraction of transmitted electrons is $(1/a)\int_{-a/2}^{a/2} dx \;|t(x)|^2=\sum_n|t_n|^2$.

Relevant types of gratings can be considered, each characterized by simple analytical relations for the coefficients $t_n$ as follows:
\begin{itemize}
\item[({\it i})] {\it Centrosymmetric grating}.---Setting $t(x)=t(-x)$, the beam amplitudes reduce to $t_n=t_{-n}=(2/a)\int_0^{a/2} dx \;t(x)\,\cos(ngx)$, and the maximum fraction of incident electrons transmitted in the $n=\pm1$ beams is $8/\pi^2\approx81$\%, as obtained, for example, with a gray-scale binary grating defined by a transmission function $t(x)=2\Theta(a/4-|x|)-1$ inside the first period ($|x|$<a/2). With this choice, even-$n$ beams are suppressed, including $n=0$.
\item[({\it ii})] {\it Two-beam grating}.---If we are interested in having only two transmitted beams (e.g., $n=\pm1$) of equal amplitude ($t_1=t_{-1}$), the transmission function can be set to $t(x)=2t_1\cos(gx)$ up to a global phase, according to Eq.~(\ref{gratingeq1}). From the physical condition $|t(x)|\le1$, the maximum beam intensity is limited to $|t_1|^2=1/4$, and consequently, only half of the electrons are transmitted. In addition, a nonbinary grating is required to accommodate the smooth $t(x)$ profile, whose fabrication can be challenging.
\item[({\it iii})] {\it Black-and-white binary grating}.---A simpler solution is provided by a slit array patterned in a film thick enough to be opaque to electrons (i.e., $t(x)\in\{0,1\}$). In this configuration, the fraction of transmitted electrons coincides with the percentage of the area covered by the slit apertures. A special case corresponds to the choice $t(x)=\Theta(a/4-|x|)$ in the first period (i.e., a slit width equal to half of the period), which suppresses all even-$n$ beams except $n=0$, while the nonzero transmission amplitudes are $t_0=1/2$ and $t_n=(-1)^{(n-1)/2}/n\pi$ for odd $n$; only half of the electrons are transmitted, with half of them in the direct $n=0$ beam and a fraction of $4/\pi^2\approx40\%$ in the $n=\pm1$ beams.
\end{itemize}
In a physical realization of gratings of types ({\it i}) and ({\it ii}), a phase $eV_0d(x)/\hbar v$ can be introduced in $t(x)$ upon transmission through a metallic film with an $x$-dependent thickness $d(x)$ and a negative inner potential $-V_0$. Considering common values of $V_0\sim10$~eV, a reasonable thickness of $34$~nm (43~nm) produces a phase shift of $\pi$ (i.e., $t(x)=-1$) for 100~keV (200~keV) electrons. Unfortunately, some decoherence and loss of probability (i.e., $|t(x)|<1$) are unavoidable consequences of traversing the material. Electrons transmitted through gratings of type ({\it iii}) should be less affected by such effects, as they do not traverse any material.

\renewcommand{\theequation}{E\arabic{equation}} 
\renewcommand{\thesection}{E} 
\section{Grating coordination}
\label{conjugation}

In our SFES design, the lateral wave vector transfer in each grating $j=1,2$ is $n\gb_j$, where $\gb_j=(2\pi/a_j)\,\xx$ (see Fig.~1a). Grating 1 transforms an incident wave with transverse wave vector $\Qb_i$ into diffracted waves with lateral wave vectors $\Qb_i+n'\gb_1$, where $n'$ runs over diffraction orders. Each diffracted wave $n'$ is then transmitted through the two central lenses and hence transformed into a wave with transverse wave vector $-(\Qb_i+n'\gb_1)(f_1/f_2)$ [see Eq.~(\ref{twolenseseq})]. Finally, grating 2 adds a lateral wave vector $n\gb_2$ with a new diffraction order $n$, thus resulting in a wave with total transverse wave vector $-(\Qb_i+n'\gb_1)(f_1/f_2)+n\gb_2$. To guarantee that the waves emerging from grating 1 are superimposed along shared diffracted waves after grating 2, we assume the geometrical condition
\begin{align}
(f_1/f_2)\gb_1=\gb_2, \label{conjugationeq}
\end{align}
or equivalently, $a_1/a_2=f_1/f_2$ [i.e., Eq.~(\ref{gratingconjugation}) in the main text].

\renewcommand{\theequation}{F\arabic{equation}} 
\renewcommand{\thesection}{F} 
\section{Grating displacement}
\label{displacement}

As a central ingredient in the design of SFES, we introduce a tunable lateral displacement $D$ in grating 2 (period $a_2$) in Fig.~1a. The transmission function [$t_2(x)$ without displacement] is then changed to $t_2(x-D)$, whose Fourier transform [Eq.~(\ref{gratingeq2})] is modified by an additional $n$-dependent phase $-ng_2D$. The transmission coefficients of grating 2 are thus transformed according to the rule
\begin{align}
t_{2,n} \rightarrow t_{2,n} \;\ee^{-\ii ng_2D} \label{displacementeq}
\end{align}
when the grating is laterally displaced by a distance $D$ along $x$.

\renewcommand{\theequation}{G\arabic{equation}} 
\renewcommand{\thesection}{G} 
\section{Electron plane-wave transmission through the SFES setup}
\label{fullSFES}

We now combine Eqs.~(\ref{eq2}) and (\ref{twolenseseq})--(\ref{displacementeq}) to describe the transmission properties of the SFES setup in Fig.~1a. In the absence of a specimen, a near-axis plane wave $\ee^{\ii\Qb_i\cdot\Rb}$ incident on grating 1 (i.e., with small $Q_i$ compared with $2\pi/a_j$ for both gratings $j=1,2$) is transformed into a wave function
\begin{align}
\phi_{\Qb_i}(\rb)=-\frac{f_1}{f_2} \sum_{n} \ee^{-\ii[(f_1/f_2)\Qb_i-n\gb_2]\cdot\Rb} \sum_{n'} t_{1,n'} \;t_{2,n'+n}
\;\ee^{-\ii(n'+n)g_2D} \;\ee^{\ii\chi_{in'}} \label{SFES0}
\end{align}
right after exiting grating 2. Here, we define the phase
\begin{align}
\chi_{in'}=\frac{|\Qb_i+n'\gb_1|^2}{2q_0}\big(f_1+f_1^2/f_2-d_1-d_2f_1^2/f_2^2\big);
\label{chiin}
\end{align}
$t_{j,n}$ are grating transmission coefficients [see Eqs.~(\ref{gratingeq})]; the two gratings are coordinated according to Eq.~(\ref{conjugationeq}); a lateral displacement $D$ is introduced in grating 2 through the substitution in Eq.~(\ref{displacementeq}); we use the transmission properties of the two-lens system as given by Eq.~(\ref{twolenseseq}); and free propagation in the spacings $d_1$ and $d_2$ separating the gratings from the lenses is accounted for through Eq.~(\ref{eq2}) with $\Qb_i$ substituted by the lateral wave vectors in the respective regions, $\Qb_i+n'\gb_1$ and $-(f_1/f_2)[\Qb_i+n'\gb_1]$.

We can trivially set $\chi_{in}=0$ by adjusting the focal and propagation distances $f_j$ and $d_j$ in Eq.~(\ref{chiin}) in such a way that
\begin{align}
\frac{f_1^2}{f_2^2}=-\frac{d_1-f_1}{d_2-f_2}, \label{lenseq}
\end{align}
which corresponds to the condition that each grating is at the image plane of the other one through the two-lens system (i.e., there is no extra phase associated with tilting of the incident e-beam away from the on-axis direction under the assumption of paraxial wave propagation).

\renewcommand{\theequation}{H\arabic{equation}} 
\renewcommand{\thesection}{H} 
\section{Gaussian e-beam transmission through the SFES setup without a specimen}
\label{Gaussianspecimen}

Electron plane waves form a suitable basis for understanding electron optics, but they are not ideal for studying systems composed of finite-size lenses and gratings. Gaussian e-beams provide a more realistic description of electron microscopes. Therefore, we consider an incident Gaussian paraxial beam expanded in terms of plane waves as
\begin{align}
\phi^{\rm inc}(\rb)&=\frac{w_0}{2\pi\sqrt{\pi}}\int d^2\Qb_i \;\ee^{-Q_i^2w_0^2/2}
\;\ee^{\ii\Qb_i\cdot\Rb} \;\ee^{-\ii Q_i^2z/2q_0} \nonumber\\
&=\frac{1}{\sqrt{\pi}\zeta w_0}\ee^{-R^2/2\zeta w_0^2}, \label{gaussianeq}
\end{align}
where $\zeta=1+\ii z/q_0w_0^2$, $w_0$ is the waist radius at $1/e$ of the maximum probability density, and we are assuming e-beams with a high degree of lateral coherence. This wave function is normalized such that $\int d^2\Rb\,|\phi(\rb)|^2=1$ for all values of $z$.

Under typical conditions in electron microscopes, the e-beam width is small compared to the diameters of the lenses and gratings. Therefore, the results derived for plane waves in the preceding sections can be used to produce physically meaningful analytical expressions once we integrate over $\Qb_i$. Additionally, we assume that the lenses and gratings are sufficiently wide to transmit all diffracted beams with substantial intensity.

In SFES, we consider a Gaussian wave incident on grating 1 with a large waist $w_0$ in the micron range. As a result, the variation of $\zeta$ with $z$ in Eq.~(\ref{gaussianeq}) is negligible for the beam distances involved in electron microscopes because $|z|\ll q_0w_0^2$ ($\sim63$~m for $w_0=5\,\mu$m and 200~keV electrons). 
Thus, we can approximate Eq.~(\ref{gaussianeq}) as
\begin{align}
\phi^{\rm inc}(\rb)\approx\frac{1}{\sqrt{\pi}w_0}\ee^{-R^2/2w_0^2}. \label{gaussian2}
\end{align}
In addition, we assume that the focal and grating distances satisfy Eq.~(\ref{lenseq}), and consequently, we have $\chi_{in}=0$ and Eq.~(\ref{SFES0}) becomes
\begin{align}
\phi_{\Qb_i}(\rb)=-\frac{f_1}{f_2} \sum_{n} \ee^{-\ii[(f_1/f_2)\Qb_i-n\gb_2]\cdot\Rb} \sum_{n'} t_{1,n'} \;t_{2,n'+n}
\;\ee^{-\ii(n'+n)g_2D}. \label{SFES1}
\end{align}
Even if Eq.~(\ref{lenseq}) is not imposed, we find $\chi_{in}\ll1$ when considering feasible submicron grating periods $a_j$; indeed, rewriting Eq.~(\ref{chiin}) as $\chi_{in}=A_0+A_1n+A_2n^2$ and taking again $w_0=5\,\mu$m (i.e., $Q_i\lesssim2/w_0\approx0.4/\mu$m) and 200~keV electrons in combination with reasonable parameter choices for $a_j$ ($\sim1\,\mu$m) and $d_j,f_j$ ($\sim$ millimeters), we obtain $A_0\sim10^{-5}$, $A_1\sim10^{-3}$, and $A_2\sim10^{-2}$. 

We now modify Eq.~(\ref{SFES1}) to describe an incident Gaussian e-beam by performing a linear average over the amplitudes contributed by each $\Qb_i$ component. More precisely, we multiply the wave function in Eq.~(\ref{SFES1}) by a factor $(w_0/2\pi\sqrt{\pi})\;\ee^{-Q_i^2w_0^2/2}$ [coming from the plane-wave decomposition of the incident wave in Eq.~(\ref{gaussianeq})], as well as an extra factor $\ee^{-\ii|(f_1/f_2)\Qb_i-n\gb_2|^2z/2q_0}$ to account for propagation along a distance $z$ measured from the position of grating 2 [see Eq.~(\ref{eq2})]. Integrating over $\Qb_i$ and noticing that $zf_1/q_0w_0^2f_2\ll1$, the transmitted wave function becomes
\begin{align}
\phi(\rb)\approx-\frac{1}{\sqrt{\pi}w_0}\frac{f_1}{f_2} \sum_{n} \ee^{-(f_1/f_2)^2|\Rb-n\gb_2z/q_0|^2/2w_0^2} \ee^{\ii n\gb_2\cdot\Rb-\ii n^2g_2^2z/2q_0} \sum_{n'} t_{1,n'} \;t_{2,n'+n}
\;\ee^{-\ii(n'+n)g_2D}, \label{SFES2}
\end{align}
which is a superposition of Gaussian beams centered around the transverse wave vectors $n\gb_2$. Each of these beams is normalized in the same way as the incident wave [cf. $(f_2/\sqrt{\pi}w_0f_1)\ee^{-(f_1/f_2)^2|\Rb-n\gb_2z/q_0|^2/2w_0^2}$ and Eq.~(\ref{gaussian2})].

We can straightforwardly write the probability that the incident electron is transmitted through the two lenses and gratings as
\begin{align}
P_0(D)=\sum_{n}\bigg|\sum_{n'} t_{1,n'} \;t_{2,n'+n} \;\ee^{-\ii n'g_2D}\bigg|^2, \label{SFES3}
\end{align}
where the subindex $0$ indicates the absence of a specimen. In Eq.~(\ref{SFES3}), the internal sum over $n'$ accounts for the coherent superposition of beams diffracted by grating 1 upon transmission through grating 2, while an incoherent sum is performed over transmitted beams $n$ [see also Eq.~(\ref{SFES2})]. We now use Eq.~(\ref{gratingeq2}) to rewrite the transmission coefficients as a function of the transmission functions $t_j(x)$. After some algebra, Eq.~(\ref{SFES3}) becomes (see Appendix\ \ref{appendixA})
\begin{align}
P_0(D)=\int_{-1/2}^{1/2}d\theta\;\big|t_1(\theta a_1)t_2(-\theta a_2-D)\big|^2 \label{SFES4}
\end{align}
[i.e., Eq.~(\ref{SFES4}) in the main text], which is a periodic function of $D$ (with period $a_2$) yielding the fraction of the overlapped area between {\it openings} in the two gratings. Interestingly, Eq.~(\ref{SFES4}) shows that the overall transmission can be modulated by varying $D$, and even suppressed if there is no overlap between openings in the two gratings.

In the tutorial configuration considered in Fig.~\ref{Fig1} of the main text, two-beam gratings (with $\pm\gb_j$ wave vector transfers) like those of type ({\it ii}) are employed [see Sec.~\ref{grating}], and only the $n=0$ on-axis beam is detected after transmission through grating 2. Assuming that the transmission coefficients $t_j=t_{j,\pm1}$ take the same value for both beams ($\pm1$) in each grating, the probability of measuring an electron is given by the $n=0$ term in Eq.~(\ref{SFES3}), which reduces to
\begin{align}
P_0(D)=4\,|t_1t_2|^2\,\cos^2(2\pi D/a_2) \nonumber
\end{align}
[i.e., Eq.~(\ref{P0}) in the main text].

In the configuration of Fig.~\ref{Fig2} in the main text, we use gratings of type ({\it iii}) [see Sec.~\ref{grating}]. When the slit apertures cover half of the surface area, the overall transmission probability, which has the periodicity $P_0(D)=P_0(D+a_2)$, becomes $P_0(D)=1/2-|D|/a_2$ in the first period (i.e., for $|D|<a_2/2$) and vanishes if $D=D_0=(m+1/2)a_2$ for any integer $m$. By making the slit openings narrower, Eq.~(\ref{SFES4}) predicts a complete suppression of transmission occurring over a finite range of $D$ values, which should render SFES more tolerant to an imperfect lateral coherence of the e-beam. For example, for a slit width of 0.4 times the period, only $40\%$ of electrons are transmitted through the grating, with 40\% of them in the direct $n=0$ beam and $\approx46\%$ in the sum of the $n=\pm1$ beams.

\renewcommand{\theequation}{I\arabic{equation}} 
\renewcommand{\thesection}{I} 
\section{Focused beam profiles}
\label{focalspots}

In the studied SFES configuration, the electron is prepared as a superposition of laterally displaced beams corresponding to the orders $n$ of diffraction by grating 1. In the focal plane $z=z_f$ between the two lenses of Fig.~1a, the beam profile for each $n$ can be obtained by convoluting Eq.~(\ref{PSF}) with the distribution of incident transverse wave vectors $(w_0/2\pi\sqrt{\pi})\;\ee^{-Q_i^2w_0^2/2}$ inherited from Eq.~(\ref{gaussianeq}) and their corresponding focal positions $\Rb_i=(\Qb_i+n\gb_1)f_1/q_0$ (i.e., including the lattice wave vector $n\gb_1$ after transmission through grating 1). We obtain
\begin{align}
\phi(\Rb,z_f)\approx-\ii\;\ee^{\ii q_0R^2/2f_1}
\sum_{n} t_{1,n} \;\dfrac{\ee^{-|\Rb-\Rb_n|^2/2w_f^2}}{\sqrt{\pi}w_f}, \label{focalplane}
\end{align}
where $w_f=f_1/q_0w_0=\lambda_e\times f_1/2\pi w_0$ is the Gaussian half-width at the foci, $\Rb_n=n\gb_1f_1/q_0=n\lambda_e\times f_1/a_1\,\xx$ is the focal position of beam $n$, and $\lambda_e=2\pi/q_0$ is the electron wavelength. The incident beam width $w_0$ is now changed to $w_f$ [cf. Eqs.~(\ref{gaussian2}) and (\ref{focalplane})]. The electron wave is thus transformed into a periodic linear array (along $x$) of focused beams with a large period $\lambda_e f_1/a_1$ compared with the focus size $\lambda_e f_1/2\pi w_0$, provided the e-beam width $w_0$ covers many periods $a_1$ upon incidence on grating 1.

\renewcommand{\theequation}{J\arabic{equation}} 
\renewcommand{\thesection}{J} 
\section{Gaussian e-beam transmission through the SFES setup with an illuminated specimen}
\label{Gaussian}

By placing an illuminated specimen in the focal plane of the SFES scheme in Fig.~1a, the focused beams described in Sec.~\ref{focalspots} are exposed to optical near fields that transform them into energy combs. We assume that the optical field varies negligibly over the width $w_f$ of each focused beam and further adopt the nonrecoil approximation, such that the transverse component of the electron wave function is not altered by the interaction with light. Before the interaction, each focused beam is characterized by a longitudinal state $\ket{0}$ of kinetic energy $\hbar\varepsilon_0$ inherited from the incident e-beam. For illumination with monochromatic light of large photon energy $\hbar\omega$ compared with the energy width of the incident electron, the longitudinal state is multiplexed into a well-defined coherent energy comb $\sum_\ell\alpha_{n,\ell}\ket{\ell}$, comprising states $\ket{\ell}$ of energies $\hbar(\varepsilon_0+\ell\omega)$ weighted by complex-valued coefficients $\alpha_{n,\ell}=J_\ell(2|\beta_n|)\;\ee^{\ii\ell{\rm arg}\{-\beta_n\}}$. These coefficients are determined by the dimensionless coupling parameters $\beta_n=(e/\hbar\omega)\int dz\,E_z(\Rb_n,z)\,\ee^{-\omega z/v}$ (i.e., the spatial Fourier transform of the optical near field $E_z(\rb)$ along the electron path \cite{paper151,PLZ10,paper371}, where we use the convention $\Eb(\rb,t)=\Eb(\rb)\,\ee^{-\ii\omega t}+{\rm c.c.}$). We note that the electron probability in each beam $n$ remains unchanged by the interaction with light, as confirmed by the identity $\sum_\ell|\alpha_{n,\ell}|^2=\sum_\ell J_\ell^2(2|\beta_n|)=1$.

Elastic and inelastic scattering by spontaneous interaction with the specimen is expected to affect the transmitted beam amplitudes. This effect could be approximately incorporated through the transmission coefficients $\alpha_{n,\ell}$, but we ignore them for simplicity in this work under the assumption that the electrons do not traverse thick material regions.

We assume that lens 2 and grating 2 produce transverse wave functions roughly independent of $\ell$, and only the longitudinal part is affected by energy multiplexing. The transmission probability in Eq.~(\ref{SFES3}) can then be separately applied to each $\ell$ component to describe the detection of an incoherent superposition of different energy sidebands (i.e., all electrons are collected without resolving them in energy). More precisely, the overall electron probability becomes
\begin{align}
P(D)=\sum_{n,\ell}\bigg|\sum_{n'} \alpha_{n',\ell} \;t_{1,n'} \;t_{2,n'+n} \;\ee^{-\ii n'g_2D}\bigg|^2 \label{SFES5}
\end{align}
[i.e., Eq.~(\ref{SFES5}) in the main text], where we have introduced the energy-multiplexing coefficients $\alpha_{n',\ell}$ associated with each focused beam $n'$. It should be noted that different $n'$ beams diffracted by grating 1 are still interfering within each final electron energy subspace $\ell$.

A simple interesting result can be derived by assuming that only one diffracted beam interacts with the optical field. We choose $n'=0$ as the interacting beam so that the rest of them (i.e., $n'\neq0$) have interaction coefficients $\alpha_{n',\ell}=\delta_{\ell,0}$. Further setting the grating configuration (i.e., their geometry and displacement $D=D_0$) such that electron transmission is suppressed in the absence of the specimen [i.e., $P_0(D_0)=0$], Eq.~(\ref{SFES5}) reduces to (see Appendix\ \ref{appendixB})
\begin{align}
P(D_0)=2\,|t_{10}|^2 \;T_2 \;\big[1-J_0(2|\beta_0|)\big] \label{SFES6}
\end{align}
[i.e., Eq.~(\ref{SFES6}) in the main text], where $T_2=\sum_n|t_{2,n}|^2$ is the transmittance of grating 2. The electron transmission probability is thus dependent on the depletion in the amplitude of the elastic $n'=0$ focused beam. For weak electron--light interaction ($|\beta_0|\ll1$), Eq.~(\ref{SFES6}) becomes $P(D_0)=2|t_{10}|^2T_2\;|\beta_0|^2$, which is quadratic in the optical field amplitude.

\renewcommand{\theequation}{K\arabic{equation}} 
\renewcommand{\thesection}{K} 
\section{Numerical calculation of the $\beta_n$ coefficients and EELS/CL probabilities}
\label{simulations}

The electron energy-loss spectroscopy (EELS) and cathodoluminescence (CL) probabilities in Fig.~\ref{Fig1} of the main text are calculated analytically from a multipolar expansion of the response of the sphere, as detailed in Ref.~\cite{paper021}, while those in Figs.~3 are obtained numerically using the boundary-element method (BEM) for axially symmetric structures \cite{paper040}.

The electron--light coupling coefficients $\beta_n$ are derived from their relationship to the CL far-field amplitude $\fb_n$. More precisely, as demonstrated in Ref.~\cite{paper371}, these two quantities rigorously satisfy the equation 
\begin{align}
\beta_n=\ii c^2 \fb_n\cdot\Eb^{\rm ext}/(\hbar\omega^2), \label{betavsf}
\end{align}
where $\beta_n$ denotes the coupling coefficient for an electron moving along a trajectory $\Rb_n+\vb t$ when the specimen is illuminated by an external plane wave with electric field amplitude $\Eb^{\rm ext}$, frequency $\omega$, and wave vector $\kb=k\,\kk$ with $k=\omega/c$ [i.e., $\Eb^{\rm ext}(\rb,t)=\Eb^{\rm ext}\,\ee^{-\ii\omega t}+{\rm c.c.}$]; the vector $\fb_n$ corresponds to the emission amplitude at frequency $\omega$ along the direction $-\kb$ produced by an electron moving along the trajectory $\Rb_n-\vb t$ [i.e., the CL electric far-field amplitude is $\Eb^{\rm CL}(\rb,t)=\fb_n\,\ee^{\ii(kr-\kb\cdot\rb-\omega t)}/r+{\rm c.c.}$, with the origin chosen near the specimen]; $\Rb_n$ is the transverse position of the electron beam; and $\vb=v\,\zz$ is the electron velocity vector. Consequently, Eq.~(\ref{betavsf}) allows us to calculate the electron--light coupling coefficients from the CL far-field amplitudes, which are, in turn, computed using the methods mentioned above, but with reversed directions of electron and light propagation.

\renewcommand{\theequation}{L\arabic{equation}} 
\renewcommand{\thesection}{L} 
\section{Derivation of equation~(\ref{SFES4})}
\label{appendixA}
\setcounter{equation}{0}

We start from Eq.~(\ref{SFES3}), where we rewrite the coefficients $t_{j,n}$ in terms of $t_j(x)$ using Eq.~(\ref{gratingeq2}). This leads to
\begin{align}
P(D)=\sum_{n}\bigg|\frac{1}{a_1a_2}
\int_{-a_1/2}^{a_1/2} dx
\int_{-a_2/2}^{a_2/2} dx'
\;t_1(x)\;t_2(x') \;\ee^{-\ii ng_2x'}
\sum_{n'}\ee^{-\ii n'(g_1x+g_2x'+g_2D)}\bigg|^2, \nonumber
\end{align}
where $a_j$ are the grating periods and $g_j=2\pi/a_j$ the reciprocal unit vectors. The $n'$ sum can be performed by applying the identity
\begin{align}
\sum_{n}\ee^{\pm\ii n\alpha}=\sum_m\delta(\alpha/2\pi-m) \label{sumn}
\end{align}
with $\alpha=g_1x+g_2x'+g_2D$, so the probability becomes
\begin{align}
P(D)=\sum_{n}\bigg|\frac{1}{a_1}
\int_{-a_1/2}^{a_1/2} dx
\;t_1(x)\;t_2(-xa_2/a_1-D) \;\ee^{2\pi\ii n(x+D)/a_1}
\bigg|^2, \label{SFESap1}
\end{align}
where we have used the fact that only one integer value $m$ survives within the range of $x$ and $x'$ integration. We have also employed the periodicity property $t_2(x)=t(x+ma_2)$. In Eq.~(\ref{SFESap1}), the squared modulus can readily be expanded to yield
\begin{align}
P(D)=\frac{1}{a_1^2}
\int_{-a_1/2}^{a_1/2} dx
\int_{-a_1/2}^{a_1/2} dx'
\;t_1(x)\;t_2(-xa_2/a_1-D)
\;t_1^*(x')\;t_2^*(-x'a_2/a_1-D)
\sum_{n}\ee^{2\pi\ii n(x-x')/a_1}. \nonumber
\end{align}
Finally, carrying out the $n$ sum by applying Eq.~(\ref{sumn}) again, we find
\begin{align}
P(D)=\frac{1}{a_1}
\int_{-a_1/2}^{a_1/2} dx
\;\big|t_1(x)\;t_2(-xa_2/a_1-D)\big|^2, \nonumber
\end{align}
which becomes Eq.~(\ref{SFES4}) by changing the integration variable to $\theta=x/a_1$.

\renewcommand{\theequation}{M\arabic{equation}} 
\renewcommand{\thesection}{M} 
\section{Derivation of equation~(\ref{SFES6})}
\label{appendixB}
\setcounter{equation}{0}

Setting $\alpha_{n',\ell}=\delta_{\ell,0}$ for $n'\neq0$ in Eq.~(\ref{SFES5}), we can directly write it as
\begin{align}
P(D)&=\sum_{n}\bigg|\sum_{n'} \alpha_{n',0} \;t_{1,n'} \;t_{2,n'+n} \;\ee^{-\ii n'g_2D}\bigg|^2
+|t_{1,0}|^2 \sum_n |t_{2,n}|^2 \sum_{\ell\neq0}|\alpha_{0,\ell}|^2, \nonumber\\
&=\sum_{n}\bigg|(\alpha_{0,0}-1) \;t_{1,0} \;t_{2,n}+\sum_{n'} \;t_{1,n'} \;t_{2,n'+n} \;\ee^{-\ii n'g_2D}\bigg|^2
+|t_{1,0}|^2 \;T_2 \;\big(1-\alpha_{0,0}^2\big), \label{PPP}
\end{align}
where the first term on the right-hand side corresponds to elastically transmitted electrons ($\ell=0$), while the second one accounts for $\ell\neq0$ components in the $n'=0$ beam. The second line in Eq.~(\ref{PPP}) is derived by noticing that $\alpha_{0,0}$ is real, using the identity $\sum_\ell|\alpha_{n,\ell}|^2=1$, and defining the overall transmission of grating 2 as $T_2=\sum_n |t_{2,n}|^2$.
We now assume that the SFES setup is prepared such that electron transmission is suppressed in the absence of a specimen (i.e., $P_0=0$). This means that each $n$ term must be zero in Eq.~(\ref{SFES3}), which implies that the $n'$ sum also vanishes in Eq.~(\ref{PPP}). After some straightforward algebra, Eq.~(\ref{PPP}) reduces to
\begin{align}
P(D_0)=2\,|t_{1,0}|^2 \;T_2 \;(1-\alpha_{0,0}), \nonumber
\end{align}
which becomes Eq.~(\ref{SFES6}) by setting $\alpha_{0,0}=J_0(2|\beta_0|)$.

\end{widetext}


%

\end{document}